# An Assessment of the CO2 Emission Reduction Potential of Residential Load Management in Developing and Developed Countries

**Extended Abstract**
Version 03-Apr-2025

Alona Zharova[1] and Felix Creutzig[2,3]

[1] Humboldt-Universität zu Berlin, Berlin, Germany
[2] Potsdam Institute for Climate Impact Research (PIK), Member of the Leibniz Association, Potsdam and Berlin, Germany
[3] Bennett Chair of Innovation and Policy Acceleration, University of Sussex, Falmer, Brighton, UK

Intermittent renewable energies are increasingly dominating electricity grids and are forecasted to be the main force driving out fossil fuels from the grid in most major economies until 2040. However, grids based on intermittent renewables are challenged by diurnal and seasonal mismatch between supply of sun and wind and demand for electricity, including for heat pumps and electric two and four wheelers. Load management and demand response measures promise to adjust for this mismatch, utilizing information- and price-based approaches to steer demand towards times with high supply of intermittent renewables. Here, we systematically review the literature estimating CO2 savings from residential load management in developing and developed nations. We find that load management holds high potential, locally differentiated with energy mix (including the respective share of renewables and fossils), climate zone, and the regulatory environment and price mechanism. Most identified studies suggest a mitigation potential between 1 and 20%. Load management becomes more relevant with higher shares of intermittent renewables, and when electricity prices are high. Importantly, load management aligns consumers' financial incentives with climate change mitigation, thus rendering accompanying strategies politically feasible. We summarize key regulatory steps to facilitate load management in economies and to realize relevant consumer surplus and mitigation potential.

## Introduction

Meeting ambitious climate targets like the Paris Agreement's goals requires not only cleaner energy supply but also smarter energy demand. In the residential sector, electricity load management – shifting and smoothing household power use through measures such as demand response and energy efficiency – has emerged as a crucial strategy for $CO_2$ reduction. Aligning home electricity demand with periods of abundant renewable generation can cut reliance on fossil-fueled peaking plants. Recent IPCC assessments and specific integrated modelling exercises suggest that demand-side measures in buildings could reduce their carbon emissions by over 50% by 2050 relative to business-as-usual, significantly aiding national decarbonization efforts (Creutzig et al, 2022, van Heerden et al., 2025). Furthermore, by curbing extreme peaks in demand, residential load management lessens the need for new power stations and network upgrades, which in turn supports climate goals by avoiding additional infrastructure emissions (van Heerden et al., 2025). Effective load management also promises economic benefits by reducing the need for costly grid infrastructure investment. Smoothing out peak loads means utilities can defer or downsize investments in generation and transmission capacity. For example, widespread electrification of heating without demand management could greatly increase peak loads. A field trial in Great Britain found that if 20% of homes adopted electric heat pumps for heating, national peak electricity demand would rise by roughly 7.5 GW (about 14%) (Love et al., 2020). Such a surge would traditionally require extensive grid reinforcement. However, if households staggered heating usage or used smart controls, the peak impact could be moderated, avoiding expensive upgrades and enhancing grid stability. This illustrates how coordinated residential demand shifting can preempt infrastructure strains as we electrify home energy use.

Household load flexibility can additionally create consumer surplus and other co-benefits. By taking advantage of time-of-use tariffs or incentive payments, consumers can lower their electricity bills or even earn rewards for reducing consumption during peak times. Empirical evidence shows that well-designed demand-side programs often lead to net savings for participants, although the distribution of benefits can vary (White & Sintov, 2020). For instance, shifting appliance use to off-peak hours or pre-cooling a home before a peak pricing period can translate into direct cost savings. At the same time, researchers caution that not all households are equally equipped to respond; vulnerable groups might save less or risk bill increases under time-varying rates if protections are not in place (White & Sintov, 2020). To maximize consumer engagement, utilities and policymakers are increasingly experimenting with gamification – incorporating game-like elements and feedback into energy apps and programs. Gamified interventions (e.g. apps with points, competitions, or rewards for energy saving) have been shown to motivate residents to conserve energy and shift usage in a fun, competitive way (AlSkaif et al., 2018). This approach can boost participation in demand response programs by tapping into social drivers and personal goals beyond just cost savings. Over time, such engagement strategies not only reduce emissions and costs, but can also improve energy awareness and literacy among consumers.

An empowered, flexible consumer base even opens the door to gradual household energy independence. As homes adopt technologies like rooftop solar panels, battery storage, and electric vehicles, intelligent load management helps households maximize self-consumption of their own clean energy. By timing when devices charge or when appliances run, a family can use more of the electricity generated on their roof and draw less from the grid. Studies indicate that storing solar energy for use during peak hours can significantly reduce a home's reliance on utility supply (Fares & Webber, 2017). In the long run, this trend towards "prosumer" (producer-consumer) activity – with homes dynamically generating, storing, and managing their energy – enhances resilience and further cuts $CO_2$ emissions at the local level. It also dovetails with the concept of smart microgrids, where communities collectively manage loads and generation to become more self-sufficient.

The research findings and implementation challenges, however, differ between developed and developing countries. In advanced economies (e.g. the U.S. or Europe), residential load management often leverages smart grid infrastructure, automated controls, and sophisticated pricing mechanisms (Parrish et al., 2020). Key issues in these contexts include improving customer participation rates, protecting privacy, and integrating new loads like electric vehicles and heat pumps without compromising grid reliability. The heat pump trial in the UK mentioned above is one example of ensuring new low-carbon technologies can be integrated through demand flexibility (Love et al., 2020). In contrast, many developing countries face more fundamental obstacles and priorities. Electricity access and reliability are ongoing concerns, and demand-side management is sometimes limited to crude measures such as scheduled power outages (load shedding) during shortages (Usman et al., 2022). Technical and economic potential for residential load shifting certainly exists in emerging economies, but realizing it requires overcoming deficits in metering infrastructure, financing, and awareness (Usman et al., 2022). For instance, countries like India and South Africa have introduced time-based tariff pilots and efficiency programs, yet progress is slowed by infrastructural constraints and regulatory hurdles (Usman et al., 2022). Moreover, the incentives for consumers to engage can differ: where power supply is inconsistent, reliability improvements might be a bigger motivator than electricity bill savings. These contextual differences underline that strategies successful in Silicon Valley may not directly transfer to Sub-Saharan Africa without adaptation. Therefore, distinguishing the lessons learned in developed settings from those in developing ones is crucial for crafting effective, equitable load management policies worldwide.

Given the growing body of literature and the complex relationships with consumer behaviour and regulatory environment, there is a need for a systematic review of residential electricity load management in the context of $CO_2$ emissions reduction. Prior reviews have tackled pieces of the puzzle – for example, examining consumer engagement barriers (Parrish et al., 2020) – but a comprehensive synthesis is lacking. Our systematic review consolidates findings across diverse studies and regions, helping to clarify what load management strategies work best under which conditions. By evaluating the impacts on emissions, grid investments, and consumer outcomes side by side, our review identifies overall potential, and variability across nation states, climate zones, and regulatory environments. Thus we provide a comprehensive assessment of how residential load flexibility can be harnessed to meet climate targets, enhance grid efficiency, and deliver benefits to consumers.

## CO2 emission reduction potential

The estimated impact of residential load management on GHG emissions spans a very broad range across the literature. Reported values range from zero (or even slight increases) up to about 80% reduction in emissions, depending on scenario specifics (Table

1). Most studies find savings of a few percent up to around 20%, while several studies report much larger potential reductions. Notably, more than half of the studies suggest a mitigation potential exceeding 5% of baseline emissions, and about one-third report reductions above 20%.

Table 1. Quantitative modeling studies estimating the potential of load management to reduce GHG emission.

| Title and Full Reference | Quotation | GHG Reduction Estimate |
| --- | --- | --- |
| **Environment Sustainability with Smart Grid Sensor** Mahadik, et al. (2024). Frontiers in Artificial Intelligence. | "This integration can lead to a reduction in greenhouse gas emissions by up to 20% and water usage by 30%." | Up to 20% GHG reduction with smart grid sensors. |
| **Optimal Economic and Environmental Aspects in Load Management** M. Mohamed, A.M. El-Rifaie, et al. (2024). Processes. | "This corresponds to an 81% reduction in GHG emissions." | 81% GHG reduction with DSM-based load management. |
| **Benefits of demand-side response in combined gas and electricity networks** M. Qadrdan, M. Cheng, J. Wu, N. Jenkins (2017). Applied Energy. | "Carbon savings were quantified by applying DSR...congestions and the GHG emissions were alleviated significantly." | Significant carbon savings via demand-side response. |
| **The effect of price-based demand response on carbon emissions in European electricity markets** M. Fleschutz, M. Bohlayer, et al. (2021). Applied Energy. | "MEF-based load shifts reduced the mean resulting carbon emissions by 35%" | 35% mitigation potential if marginal $CO_2$ emissions are used for guiding load shifting |
| **Demand response and other demand side management techniques for district heating: A review** E. Guelpa, V. Verda (2021). Energy. | "A 9% reduction in energy costs and GHG emissions in district heating systems." | 9% reduction in GHG emissions from DSM measures. |
| **A Review of Current and Future Costs and Benefits of Demand Response for Electricity** P. Bradley, M. Leach, J. Torriti (2011). Centre for Environmental Strategy. | "GHG reductions from electricity reductions... quantified in the DECC and Ofgem 2010b report..." | At least 2,8% in terms of energy use alone... |
| **Assessment of the Cost and Environmental Impact of Residential Demand-Side Management** Tsagarakis et al (2016). IEEE Transactions on Industry Applications | "maximum GHG emissions savings as a proportion of the emissions before any DSM actions (the latter calculated from the average emissions factors) | 1.96% |
| **Carbon savings in the UK demand side response programmes** Lau et al (2015). Applied Energy | "The 132 trial days of simulation indicate the overall 1.83 ± 0.54% reduction of $CO_2$ achieved per consumer." "if there are 500,000 people in a large city, the overall $CO_2$ savings for 132 days would be 14.60 ± 0.02 $ktCO_2$" | 1.83% per consumer from DSR programmes |

| Reference | Quote | Value |
|---|---|---|
| [From peak shedding to low-carbon transitions: Customer psychological factors in demand response](#)<br>Lin et al (2022) Energy | "In the environmental aspect, the carbon emissions were reduced by more than half compared to the start point." | 50% |
| [Does demand-side flexibility reduce emissions? Exploring the social acceptability of demand management in Germany and Great Britain](#)<br>Gleue et al (2021) Energy Research & Social Science | "With 15% flexibility the carbon saving range from 1.1% to 1.5% for Germany and Great Britain respectively." "Assuming that all households who are potentially willing to accept time-variant prices actually shift 15% of their consumption, our model suggests direct emission savings of 0.8% in German and 0.9% in British society" | 1.1 - 1.5% carbon savings if willing households shift 15% of their energy consumption |
| [The role of flexible demand to enhance the integration of utility-scale photovoltaic plants in future energy scenarios: An Italian case study](#)<br>Veronese et al (2024) Renewable Energy | "In terms of RES (Renewable Energy Sources) penetration and CO2 emissions savings, the introduction of flexibility at the consumption side has no significant effects being less than 1 % in all the scenarios examined. This result suggests that the total amount of flexible demand available for load shifting shall go hand in hand with the increasing VRES (Variable Renewable Energy Sources) penetration, that is the main source for reducing the CO2 emissions" | 1% |
| [Smart Scheduling of Household Appliances to Decarbonise Domestic Energy Consumption](#)<br>Stacpoole et al (2019) IEEE/CIC International Conference on Communications Workshops in China | "... on average, a household with only grid supply can reduce the carbon footprint of a wet appliance by 23.9% by optimally scheduling its start time." "where the household has a PV supply and the appliance start time is scheduled … 74.7%." | from 23.9% to 74.7% |
| [The greenhouse gas reduction effect of critical peak pricing for industrial electricity: Evidence from 285 Chinese cities, 2003–2019](#)<br>Wang et al (2024) Energy Policy | "cities participating in PCPPIE underwent a significant 11.3% reduction in the GHGs emitted per industrial enterprise" | 11.3% GHG reduction in industry |
| [Field Experimental Evidence on the Effect of Pricing on Residential Electricity Conservation](#)<br>Burkhardt et al (2023) Management Science | "Our critical peak pricing intervention … leading to greenhouse gas emission reductions of about 16%" | 16% |
| [The (alleged) environmental and social benefits of dynamic pricing](#)<br>Harding et al (2023) Journal of Economic Behavior & Organization | "large-scale RCT …consumers, enabled with programmable communicating thermostats and notified of price increases a day in advance, shift demand from peak to off-peak hours… The impacts on emissions are varied. Depending on the source-generation mix of a region, the net effect on emissions from shifting electricity demand from peak to off-peak hours may yield net increases in emissions." | ? 0.27% of average monthly emissions |

| [Carbon-Aware Demand Response for Residential Smart Buildings](#), Zou et al. (2024), Electronics | "the integrated cost of the total annual electricity consumption of household devices is reduced by 8.69%" | 8% |
|---|---|---|
| [Optimizing Household Energy Use: An Activity-based Recommendation System for Reducing CO2 Emissions](#), Zharova et al. (2024), Wirtschaftsinformatik 2024 Proceedings | "The system provides an average of 12% of emissions savings and 7% of cost savings when focusing on CO2 emissions." | 12% |

## Key factors influencing GHG outcomes

**Carbon intensity of grid electricity mix.** The carbon intensity of peak vs. off-peak generation is a dominant factor in determining emissions benefits. In regions where peak demand is met by inefficient fossil fuel plants and off-peak by cleaner baseload or surplus renewables, shifting load away from peaks clearly reduces emissions. In other regions where off-peak electricity is more carbon-intensive, peak-shaving yields no benefit or a slight increase in $CO_2$ (Harding et al. 2023).

**Degree of load shift and reduction.** The fraction of load that can be flexibly shifted or shed also drives the outcome. Small adjustments (e.g. a few percent of total load shifted) naturally yield small emissions changes. On the other hand, more aggressive interventions covering a larger share of load (e.g. cycling major appliances, HVAC systems, or EV charging) can amplify impact. The Texas critical peak pricing experiment achieved 16% emissions reduction on the peak hours of the hottest days by curtailing roughly 14% of total electricity during peaks (Burkhardt et al. 2023). Notably, 74% of that response came from air conditioner load, which constitutes a large portion of summer peak demand.

**Timing.** How exactly the load is managed over time is critical. Peak cutting vs. load shifting have different implications – a reduction in overall consumption during peaks (with no full "rebound" later) yields direct emissions cuts proportional to the energy saved. Many demand response programs, however, involve shifting usage to other times. If shifted load is made up later the same day, total energy use might remain the same and only the timing differs. In such cases, emissions reduction comes only from the difference in generation mix between the two periods. If the "off-peak" period is not much cleaner, the benefit is minimal.

**Integration of renewables and storage**. The presence of distributed energy resources can greatly augment load management's impact. When households have rooftop solar panels or home batteries, load shifting can maximize the utilization of on-site renewables or cheap stored energy, displacing grid electricity that may be fossil-fueled.

**Automation.** Another factor is the level of automation and sophistication in the demand response controls. Manual or purely behavior-driven programs tend to have smaller and more variable effects (e.g. households might not consistently respond to price signals or might only reduce a little usage). In contrast, if smart thermostats, water heater controllers, or EV chargers automatically respond to grid signals, a larger portion of flexible load can be reliably shifted or shed with minimal inconvenience (Harding et al. 2023, Burkhardt et al. 2023). Increased automation generally leads to more load being shifted at the right times, hence more emissions abatement. However, automated responses must be designed carefully to align with carbon goals – for instance, an AI system optimizing solely for electricity cost might inadvertently increase emissions if prices don't reflect carbon intensity.

**Developing country considerations - focus on Africa.** Least developed countries require a specific focus in that electricity grids are often not fully developed, and thus load management has different characteristics, and even provides opportunities for leapfrogging. In Africa, load management - particularly in rural microgrids - is emerging as a critical strategy to improve energy access, system reliability, and economic viability while reducing reliance on diesel. In off-grid contexts, solar PV-battery microgrids face intermittency and load balancing challenges, which demand-side management (DSM) addresses through load shifting, peak clipping, and valley filling. Case studies from East Africa show peak demand reductions of 30% and system cost savings of up to 46% by aligning consumption with solar availability (Philipo et al., 2020; Gelchu et al., 2023). Importantly, DSM can enable microgrids to operate entirely on renewables by avoiding diesel generator use. Field projects confirm this, showing how strategic appliance upgrades, load scheduling, and productive-use programs (e.g. refrigeration businesses or health clinics adopting efficient devices) enhance both grid utilization and community welfare (Energy 4 Impact & Inensus, 2019).

On national grids, similar principles apply. South Africa's longstanding DSM initiatives - such as mass lighting retrofits and ripple-controlled water heaters - have reduced peak loads by several gigawatts, helping to defer coal-based generation and reduce $CO_2$ emissions (Usman et al., 2022). More recently, "smart load limiting" pilots allow for equitable power rationing during shortages, preserving essential services while managing demand in real time (Matsuda-Dunn et al., 2024). In Nigeria, prepaid metering and education campaigns are piloting behavioral load shifting to flatten residential peaks. Across both off-grid and grid-connected systems, the evidence is clear: DSM not only enhances technical and economic performance but also delivers measurable emissions reductions by displacing diesel and improving renewable utilization. Load management, therefore, is a linchpin in Africa's sustainable electrification strategy - maximizing the benefits of limited generation while advancing social, economic, and environmental objectives.

## Conclusion

Residential load management presents a significant opportunity for reducing $CO_2$ emissions across both developed and developing countries. Our review reveals that, while estimates vary widely, most studies suggest mitigation potentials between 1% and 20%. In developed nations, advanced metering infrastructure and dynamic pricing support load management schemes that align economic and environmental goals. In contrast, developing countries - particularly in sub-Saharan Africa - demonstrate promising potential through microgrids and targeted DSM initiatives that displace diesel reliance and enhance energy access.